\documentclass[twocolumn,showpacs,preprintnumbers,amsmath,amssymb]{revtex4}
\usepackage{graphicx}
\usepackage{dcolumn}
\usepackage{bm}
\usepackage{tensor}

\begin{document}
\title{Phase-conjugation of the isolated optical vortex 
\\using a flat surfaces.}

\author{A.Yu.Okulov}
\email{alexey.okulov@gmail.com}
\homepage{https://sites.google.com/site/okulovalexey}
\affiliation{Russian Academy of Sciences, 119991, Moscow, Russia}
\date{\ August 14, 2014}
 
\begin{abstract}
{The robust method for obtaining the helical interference 
pattern due to the phase-conjugation of an isolated optical 
vortex by means of the non-holographic technique is proposed. 
It is shown that a perfect wavefront-reversal of the 
vortex in a linear 
polarization state via even number of reflections is achievable 
due to the flip-flop of the photon's momentum 
$\vec p \approx \hbar \vec k$ with respect to the 
photon's orbital angular momentum projection $L_z$. 
The possible experimental realization is based 
upon $catseye-prism$ like reflections inside the 
confocal optical $loop$ cavity. The alternative scheme 
contains the Dove prism embedded in the optical $loop$ 
with the odd number of reflections from mirrors. 
This $confocal$ interferometric technique is applicable to 
the optical tweezers, atomic traps, Sagnac laser loops 
and metamaterials fabrication.}
\end{abstract}

\pacs{42.65.Hw,42.65.Jx,42.65.Re,42.55.Wd,42.60.Jf}

\maketitle 

\section {Introduction.}

Phase-conjugation (PC) proved to be an efficient tool for 
the laser beam divergence control \cite{Hellwarth:1977}, 
self-adjustment of optical schemes 
\cite{Zeldovich:1985} and beam combination 
\cite{Basov:1980} a decades ago. A substantial 
progress in understanding of the physical mechanism 
of a PC-mirror is associated with a concept of the 
phase-singularities inside an optical speckle patterns 
\cite{Berry:1974}. 
In accordance with this concept the randomly spaced dark 
lines of the speckle (zeros of 
electric field complex amplitude $
{\bf E}_f(z,\vec r_{{\bot}},t)$) are collocated with 
the helical phase ramps \cite{Soskin:1995}. Thus 
the phase-conjugated 
replica 
${\bf E}_b(z,\vec r_{{\bot}},t)={\bf E}^{*}_f(z,\vec r_{{\bot}},t)$ 
ought to have the set of 
the own helical phase ramps collocated with the 
phase ramps of the incident 
wave \cite{Zeldovich:1985}. 
This helical phase feature 
of the optical speckle imposes a serious limitation 
upon the usage of the deformable 
adaptive mirrors because the smooth deformable surface is 
not capable to follow the helical phase ramp. 
On the other hand the dynamical interference pattern 
written by the incident speckle 
and reflected wave inside nonlinear optical medium, 
say Stimulated Brillouin Scattering(SBS) medium 
\cite{Basov:1980} or photorefractive medium  
\cite{Woerdemann:2009}, operates like a high-fidelity 
spatial filter increasing the signal-noise ratio for the 
backward reflected PC wave ${\bf E}_b$. 

Recently the concept of phase singularity had been 
enriched by understanding 
that helical phase ramps are the sources of the 
helical interference patterns around zeros of the speckle optical 
fields \cite{Okulov:2009,Okulov:2008}. 
In particular it was shown that interference of the two 
counter propagating 
isolated optical vortices in the form of Laguerre-Gaussian 
(LG) beams produces a helical optical potential or 
"lattice with twist" \cite{Bhattacharya:2007}. 
The key point for achieving such a helical 
interference pattern proved to be the conservation 
of the $total$ orbital angular momentum (OAM) in a PC - mirror:  
the turn of OAM of reflected wave 
is the urgent requirement to the 
perfect coincidence of the incident and reflected 
wavefronts and helicoidal interference 
\cite{Okulov:2008}. Noteworthy that for the non PC mirror 
the OAM is not reversed and the interference pattern around 
phase singularity is a toroidal one \cite{Rempe:2007}.
The other important feature of the PC - mirror 
is that OAM conservation leads 
unavoidably to the transfer of rotations to the 
PC mirror. In SBS mirror the rotations appear 
in the form of the helical 
acoustical phonons with $2\hbar$ OAM hence optical 
anisotropy (chirality) emerges in 
initially isotropic SBS medium \cite{Okulov:2008J}. 
Quite recently the chiral sound excitations in 
an initially isotropic liquid were found 
experimentally and obtained 
numerically using Khokhlov-Zabolotskaya-Kuznetsov equation 
\cite{Khokhlov:2008}. Nevertheless we will show below that 
in a definite experimental conditions the PC reflection 
of a single optical vortex with a topological 
charge $\ell$ may be achieved experimentally with the even number 
of reflections from the perfectly flat (nonchiral) surfaces. 
\section {Propagation of the speckle and isolated vortex line.}
\begin{figure}
\center{\includegraphics[width=0.7\linewidth] {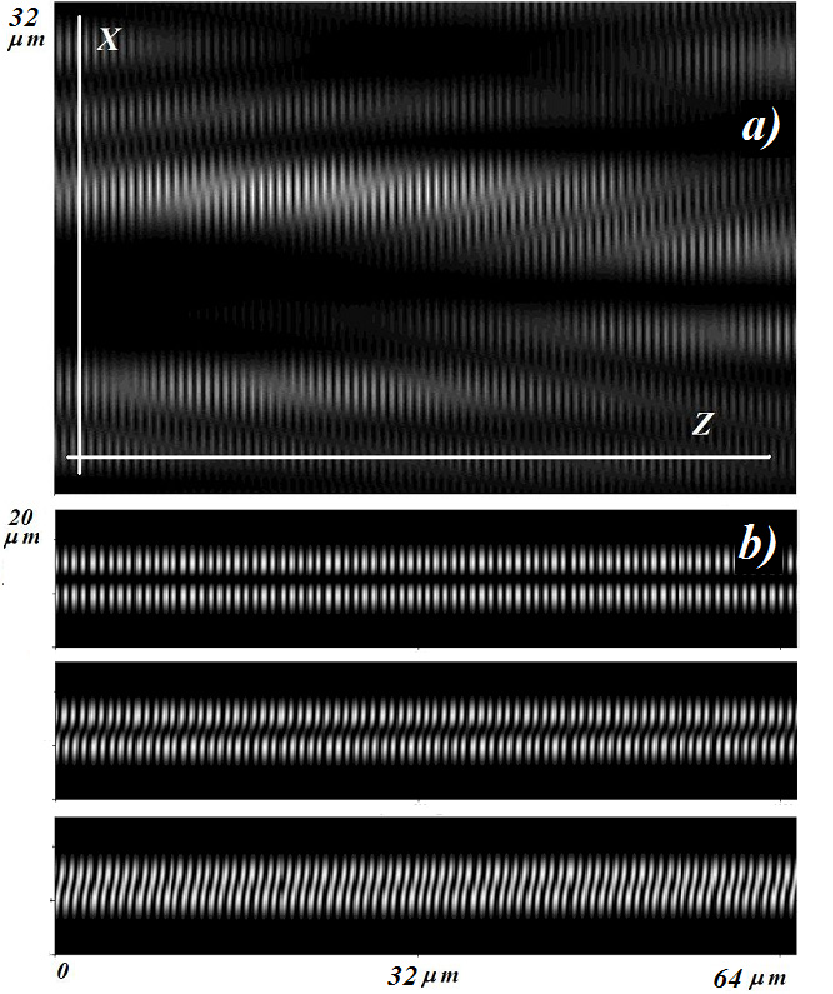}}
\caption{ a) Intertwining of the helical interference patterns 
in the optical speckle \cite{Okulov:2009}.The size of pattern is 32$\mu m$ in X-direction 
and 64$\mu m$ in Z-direction. The period of longitudinal ($Z$) modulation 
is $\lambda /2$. b) In contrast to optical vortices in a speckle 
the isolated LG optical vortices propagate rectilinearly. 
Interference pattern 
(\ref{inter_patt11}) is sliced 
at $Y=0,5,10 \mu m$ distances from the vortex axis $Z$.}
\label{fig.1}
\end{figure}

The propagation of a speckle field along $z-axis$ means a motion 
of the field zeros, i.e. the motion of the 
phase singularities in the same $z-axis$ direction.  
The trajectories of zeros are not 
rectilinear \cite{Okulov:2008J,Padgett:2005,Padgett:2009}, 
moreover trajectories intertwine each other as it happens 
with the higher-order LG optical vortices propagation 
\cite{Dholakia:2002}. The intertwining produces 
the structurally stable twisted entities in a 
speckle (fig.\ref{fig.1}) as is shown 
by numerical modeling of the following 
equation \cite{Okulov:2009}: 
\begin{equation}
\label{opposite waves}
{\frac {\partial {{{\bf E}_{(f,b)}}(z,\vec r_{{\bot}},t )}} {\partial z} }+
{\frac {n(z,\vec r_{{\bot}})}{c}}{\frac {\partial {{\bf E}_{(f,b)}}} {\partial t} }{\pm}
{\frac {i}{2 k_{(f,b)}}}{\Delta_{\bot} {{\bf E}_{(f,b)}}} = 0,
\end{equation}
where $n(z,\vec r_{{\bot}})$ is inhomogeneity of refractive index, 
$k_{(f,b)}=|\vec k_{f,b}| \approx k_z$ are the wave numbers 
of the counter directed incident and reflected 
speckle fields, with boundary condition as a multimode 
random field \cite{Okulov:1991} composed of $N_g$ plane 
waves having amplitudes $A_{{j_x} ,{j_y}}$ , random 
phases $\theta_{{j_x} ,{j_y}}$ and randomly tilted 
wave vectors each having random transverse projections 
${\vec k^{\bot}_{{j_x} ,{j_y}}}$  at $z=0$ plane: 
\begin{equation}
\label{cauchi}
\ {{\bf E}_{(f)}}(\vec{r},0 ) \approx  {{\bf E^{0} }_{(f)}} {\:}{\:} 
{\sum_{{{j_x} ,{j_y}}\in N_g}} 
 A_{{j_x} ,{j_y}}{\:}
exp{\:}[i {\:} {\vec k^{\bot}_{{j_x} ,{j_y}}} \cdot {\vec r_{\bot}}+ i \theta_{{j_x} ,{j_y}}  ].
\end{equation}
The paraxial propagation of the randomly tilted plane 
waves produces 
the twisted interference patterns resembling visually 
the $ropes$ each composed 
of several intertwined optical 
vortices \cite{Okulov:2009,Dholakia:2002}.
The similar propagation behavior and appearance of 
the $knot$ structures 
had been reported in \cite{Padgett:2005,Padgett:2009}.

\begin{figure}
\center{\includegraphics[width=0.7\linewidth] {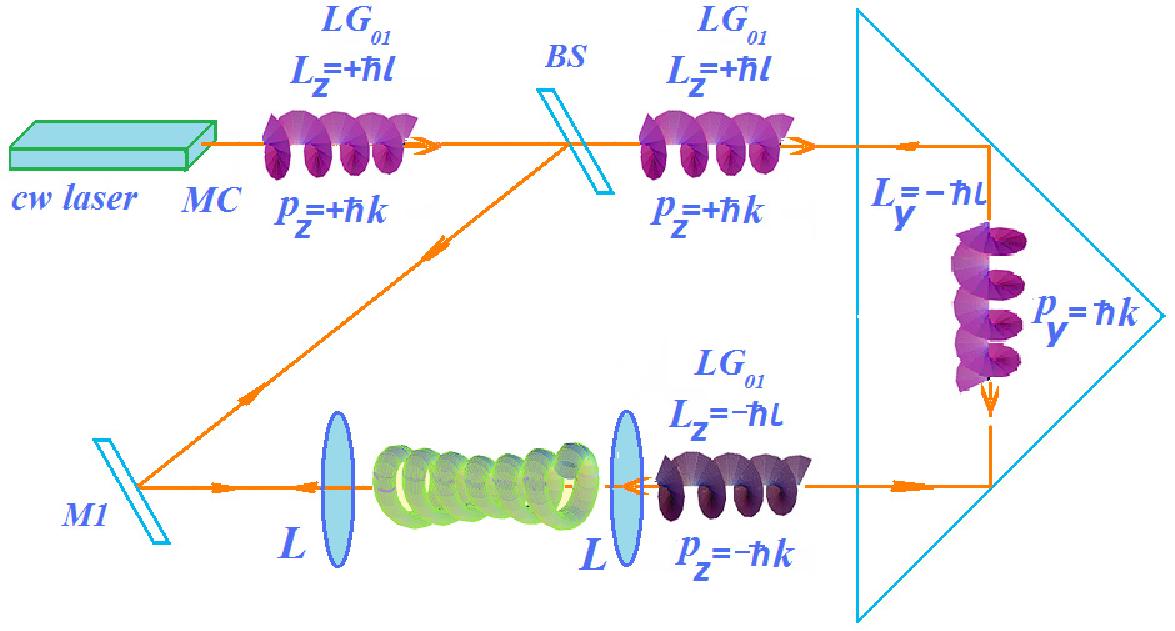}}
\caption{Two consecutive reflections of LG beam emitted 
by CW laser 
with mode converter MC. 
Each total internal reflection inside the $catseye$ prism 
from plane prism surface changes the 
topological charge of LG $\ell$ to the opposite one. After two 
reflections at 45-degree angle the optical vortex
 has the opposite direction of propagation and 
opposite direction of the angular momentum. 
The counter propagating LG 
has the same topological charge $\ell$ hence composite wavetrain 
produces the helical interference pattern. Confocal 
telescope consisting of 
the thin lenses $L$ compensates the free-space 
propagation parabolic wavefront.}
\label{fig.2}
\end{figure}
In contrast to the speckle field, the isolated vortex line 
propagates rectilinearly in a free space and it is 
structurally stable (fig.\ref{fig.1}). 
This happens for example for the 
LG laser beam with topological 
charge $\ell$ \cite{Allen:1992}: 
\begin{eqnarray}
\label{Laguerre1}
{{\bf E}_{(f,b)}(z,r,\theta,t)} 
\sim 
{\frac {{{\bf E^{0} }_{(f,b)}}{\:} exp {\:} [ {\:}i( -\omega_{(f,b)}t 
\pm k_{(f,b)} z) \pm i{\ell}\theta]} { {(1+iz/(k_{(f,b)} {D_0}^2))^2}} } {\:}
&& \nonumber \\
{\:} {({r}/{D_0})^{\ell}} {\:}
exp  {\:} [ {\:} - {\frac {r^2}{{D_0}^2(1+iz/(k_{(f,b)} {D_0}^2))}} ]{\:}{\:}{\:}, 
\end{eqnarray}
\begin{figure}
\center{\includegraphics
[width=0.7\linewidth] {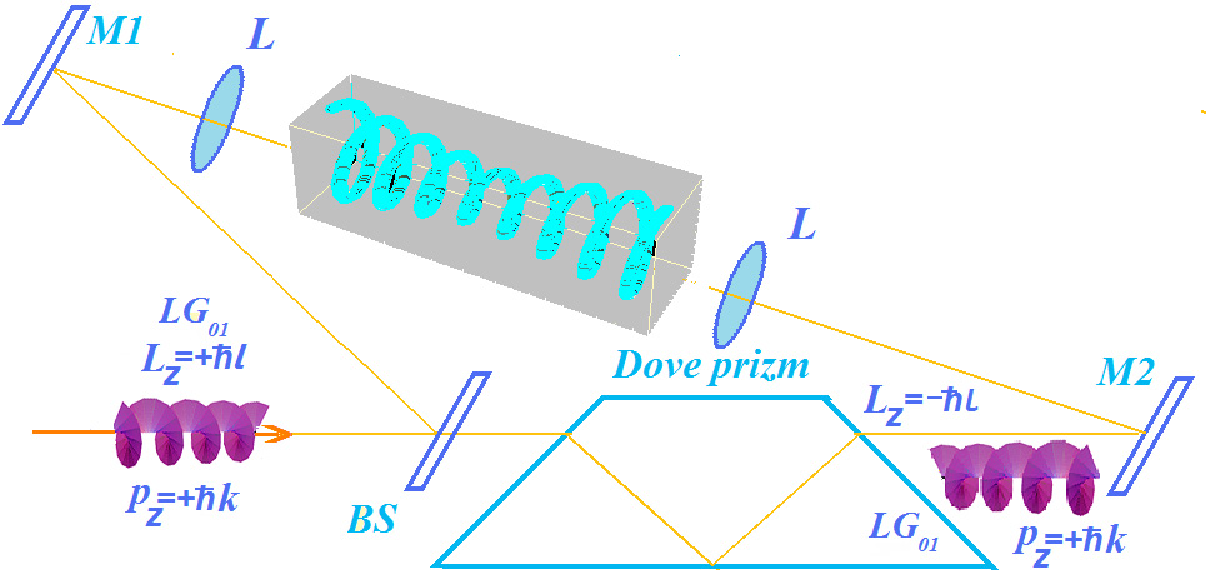}} 
\caption{The single reflection inside 
Dove prism changes the topological charge. After the 
two refractions the LG propagates with conserved momentum and overturned angular momentum. 
This leads to helical interference pattern due to overlapping 
with counter propagating LG, having opposite OAM.}
\label{fig.3}
\end{figure}
where [$(z,r,\theta,t)\rightarrow (z,\vec r_{{\bot}},t) $] are 
cylindrical coordinates embedded 
at LG axis ($z-axis$). This straight vortex line is the exact self-similar 
solution of the free space wave equation in a paraxial 
approximation (\ref{opposite waves}). 
Our aim is to describe how to use the structural 
stability, hence self-similar propagation of 
the optical vortex for the non-holographic wavefront reversal 
by conventional mirrors, lenses and prisms. 
At the first sight our proposal looks a counterintuitive 
one, because we focus attention of 
experimentalists upon previously criticized $catseye$ prism 
PC techniques \cite{Zeldovich:1985}. The case is 
that the $rays$ $reversal$ 
(with small lateral displacement) inside the prism 
is not able to perform a 
$wave$ $propagation$ reversal of the random 
collection of optical vortices 
in a speckle field or in the complicated image 
optical field. This seeming paradox is resolved 
by taking into account that rays reversal means a 
photon's $momentum$ reversal, while helical phase singularity 
is reversed by means of the $angular$ $momentum$ 
resersal \cite{Okulov:2008}. Noteworthy that 
OAM direction is changed to the opposite one inside 
Dove prism due to one total internal reflection  inside 
prism (at 45 degrees incidence angle) and two 
refractions  \cite{Barnett:2002}. The same 
happens in $catseye$ prism due to the same reason: 
when the plane surface is tilted at the angle $\alpha$ with 
respect to propagation of vortex the rotational 
symmetry of setup is absent hence the angular 
momentum is not conserved and OAM is rotated at $2 \alpha$ angle.
For both prisms the change of angular momentum is 
$2 \hbar$ per photon hence the prisms feel  
the torque $|\vec T|=2 \cdot I/ \omega$, where $I$ is 
intensity, $\omega$ is radiation 
frequency \cite{Okulov:2008}.

As a consequence of OAM reversal 
the vortex propagates in the optical 
$loop$ schemes (fig.\ref{fig.2}, fig.\ref{fig.3}) as a 
perfectly phase-conjugated one due to the simple 
reflections from 
conventional prism surfaces, provided the vortex is 
slightly focused by a thin 
lenses in order to compensate diffractive divergence. The 
technical requirements for the $loop$ adjustment are 
the same as those previously formulated for the 
ring lasers and Fabry-Perot cavities with Hermite-Gaussian 
and Laguerre-Gaussian beams \cite{Rempe:2007}.
\section{Angular momenta orientation and rotation of interference pattern}
For the ultimate quality 
PC reflection of the linearly polarized (${{\bf E^{0} }_{(f,b)}}||{\bf y}-axis$) 
$\ell$'s order LG-laser beam 
the interference pattern inside the beam waist reads as:
\begin{eqnarray}
\label{inter_patt11}
|{{\bf E}_f(z,r,\theta,t)}+{{\bf E}_b(z,r,\theta,t)}|^2 \sim 
|{{\bf E^{0} }_{(f,b)}}|^2 \cdot
&& \nonumber \\ 
{\:} {({r}/{D_0})^{2 \ell}}  exp  {\:} [ {\:} - {\frac {2 \cdot r^2}{{D_0}^2(1+iz/(k_{(f,b)} {D_0}^2))}} ]\cdot
&& \nonumber \\
{[1 + {\:}cos[  {\:} (\omega_f-\omega_b ) t - (k_f+k_b) z + 2 {\ell}{\:}\theta {\:}]]{\:}},
\end{eqnarray}

The helicity of pattern is due to the self-similar phase 
argument $(\omega_f-\omega_b ) t - (k_f+k_b) z + 2 {\ell}{\:}\theta$ 
which remains a 
constant at the double helix with a diameter $\sim 2D_0$ 
and  $\lambda /2$ pitch 
($\lambda = 2 \pi /k_{(f,b)}$) 
 \cite{Okulov:2008}. 
Such double helix optical potential rotates with angular 
frequency $\Omega=\omega_f-\omega_b $ which looks attractive from 
the point of view of optical microfluidics, micro and 
nano-particles manipulation \cite{Woerdemann:2009,Dholakia:2002} 
and as an optical dipole trap for 
ultracold atomic ensemble \cite{Bhattacharya:2007}. 

The key point in physical interpretation of this helical pattern 
is the mutual orientation of the photons 
momentum $\vec p \approx \hbar \vec k$ 
and projection of the photon's orbital angular momentum $L_z$ on 
propagation axis \cite{Okulov:2008,Allen:1992}. 
The mutual orientation 
of both quantum and classical momenta $\vec p$ and 
angular momenta $\vec L$ 
is changed after single reflection from isotropic  
optical element namely metal or multilayer dielectric mirror. 
On the contrary 
the anisotropic structures inside wavefront reversal 
mirror \cite{Okulov:2008} perform turn of 
the orbital angular momentum 
of laser beam because of the wavefront matching property of PCM. 
This turn operation is analogous to the photon's spin 
turn (change of the circular polarization 
to the oppositely rotating one) when passing through 
birefringent plate 
(i.e.anisotropic crystal) \cite{Beth:1936}.   

Consider two optical $loop$ schemes (fig.\ref{fig.2}, 
fig.\ref{fig.3}) composed 
of plane mirrors, ideal thin lenses for the adjusting 
of the parabolic component of the wavefronts (\ref{Laguerre1}) 
\cite{Barnett:2002} and 
prisms (possibly with laser gain medium inside). 
As is shown in \cite{Okulov:2008} 
each reflection from plane 
mirror changes the mutual orientation 
of the photons momentum $\hbar \vec k_z$ and the angular 
momentum $L_z= \pm \hbar \ell \vec z /z$ 
to the opposite one. 

Consequently two reflections in (fig.\ref{fig.2}) scheme 
does not change the topological charge of 
photon and oppositely propagating wave possesses the 
helical wavefront 
with the same handedness. Thus LG beam reflected inside 
$catseye$ prism 
and the other LG beam reflected from beamsplitter BS and 
mirror M1 will have the perfect wavefront coincidence 
provided their parabolic phase profiles which occurs 
due to a free-space propagation are compensated by a thin lens 
({fig.\ref{fig.2}, fig.\ref{fig.3}}). 
As a result the interference pattern 
will have a double helix geometry, provided their path 
difference is smaller than coherence 
a length $\Delta l_{coh} = c \cdot \tau$ 
($\tau$ is coherence time). Alternatively 
in fig.\ref{fig.3} scheme the 
$single$ reflection inside Dove prism 
changes the topological charge of each photon 
to the opposite 
one \cite{Barnett:2002} and the else 
reflection from mirror M2 is needed to restore the mutual 
orientation of the OAM and momentum. This sequence of 
reflections ensure the helical wavefront coincidence 
and produces 
the helical interference pattern with the twice-reflected (BS+M1) 
counter propagating LG beam. The removal of Dove prism will 
produce toroidal interference pattern because of the absence 
of phase conjugation and parallel orbital angular momenta 
of colliding photons \cite{Woerdemann:2009,Okulov:2008,Rempe:2007}.   

The frequency shift $\Omega$ may be produced via 
 two different mechanisms. The first mechanism is the rotational 
Doppler shift which arises because of rotation of 
the birefringent half-wavelength plate which alternates 
the spin component of angular 
momentum \cite{Dholakia:2002,Dholakia:2002D} 
or rotating Dove prism, which alternates the orbital component 
of angular momentum \cite{Felde:2008}. 
The Dove prism rotation technique is 
difficult to implement because of strict alignment requirements 
for interference pattern control.  
The other mechanism is the Sagnac frequency shift which 
appears in a ring laser located in rotating reference 
frame. 
This happens when prisms have 
laser gain areas collocated with LG beam propagation. 
Typically the optical gain 
is induced in a rare-earth doped dielectric host crystals 
by virtue of the diode laser pump \cite{Okulov:1993}. 
In this case the external laser outside the $loop$ 
is not necessary and the beamsplitter BS is to be replaced to 
return mirror R3. The conditions for the selection of a 
given transverse LG mode are to be fulfilled \cite{Scully:1997} 
and such a case deserves a special consideration elsewhere.
As is well known 
for the $loop$ laser schemes the counter propagating 
beams have a different frequencies $\omega_f$ and $\omega_b$ 
because of the Earth rotation having angular frequency 
$\Omega_{\oplus}$ and the angular frequency of 
the optical table rotation  
$\Omega_{_{lab}}$. For the 
such $Sagnac$ $loop$ \cite{Scully:1997} the frequency 
splitting is: 
\begin{equation}
\label{sagnac}
\Omega = (\omega_f - \omega_b) 
={\frac {8 \pi A{\:} \Omega_r}{P \cdot \lambda}}, 
{\:}{\:}{\:}{\:}
\end{equation}
where $\Omega_r = \Omega_{\oplus}+\Omega_{_{lab}} \sim (2\pi/{86400})+\Omega_{_{lab}}$ 
is the angular speed of rotation 
of the laboratory frame, $P$, $A$  are the 
perimeter and the square of the 
loop respectively. The frequency shift is measured by a 
detection of a beats (rotation of interference pattern 
in our case) of the 
counter propagating intracavity beams 
behind the cavity mirrors (M1,M2 in 
fig.\ref{fig.2}, fig.\ref{fig.3}). For the typical ratio of the 
spatial dimensions of the $Sagnac$ $loop$ $laser$ to 
the wavelength $\lambda \sim 1 \mu m$ 
the frequency splitting proves to be $\Omega \approx 
2\pi 10^{-(1-3)} rad/sec$. In particular the evaluation of 
$\Omega$ is straitforward for the circular 
ring cavity of radius $R$
when $P=2 \cdot \pi R$, $A=\pi R^2$: the frequency splitting 
is $\Omega=\Omega_r 4 \pi R /{\lambda}$.  
\section{Conclusion}
In summary we proposed the phase-conjugation 
of an isolated optical 
vortex line (LG-beam) with lateral displacement in 
the confocal optical $loop$ scheme 
with the $even$ number 
of reflections. The alternative optical loop with the $odd$ 
number of mirrors 
contains a Dove prism which alternates the 
photon's OAM projection after 
the straight passage through a prism. This scheme is different 
from Mach-Zehnder setup used previously for rotational Doppler 
effect study 
\cite{Dholakia:2002,Dholakia:2002D,Felde:2008}. 
Our loop setups with 
colliding phase-conjugated optical vortices and 
helical interference 
patterns therein are the promising tools for 
nonexpensive replacement of 
nonlinear optical phase conjugators based 
upon SBS \cite{Zeldovich:1985,Basov:1980}, photorefractive 
crystals \cite{Woerdemann:2009} and liquid crystal light valves. 
The field of experimental applications of confocal 
loops with catseye prism or Dove prism 
is in atomic traps \cite{Bhattacharya:2007} 
and optical tweezers, in particular in 
assembling the protein - like clusters \cite{Zerrouki:2008}. 
The other intriguing application is in the 
lithography of metamaterials \cite{Veselago:1968,Thiel:2010} 
and optical waveguides with the helical refractive index 
and conductivity \cite{Menachem:2006}.

\end{document}